\documentclass[aps,preprint,pre,showpacs,floatfix]{revtex4}
\usepackage{amsmath}
\usepackage[dvips]{graphicx}
\usepackage[dvips]{epsfig}
\usepackage{dcolumn}
\usepackage{bm}
\usepackage{natbib}
\usepackage{array}	
\usepackage{inputenc}

\begin{document}
\bibliographystyle{plain}

\title{Exact Results for the Kuramoto Model \\ with a Bimodal Frequency Distribution}
\author{E. A. Martens$^{1}$, E. Barreto$^{2}$, S.H. Strogatz$^{1}$, E. Ott$^{3}$, P. So$^{2}$, and T.M. Antonsen$^{3}$\\
\it
\small $^1$Department of Theoretical \& Applied Mechanics, Cornell University, Ithaca, NY 14853, USA\\
\small $^2$Department of Physics \& Astronomy and the Krasnow Institute for Advanced Study, George Mason University, Fairfax, VA 22030, USA\\
\small $^3$Institute for Research in Electronics and Applied Physics, Department of Physics, and Department
of Electrical and Computer Engineering, University of Maryland, MD 20742, USA\\
}
\affiliation{}
\begin{abstract}
We analyze a large system of globally coupled phase oscillators whose natural frequencies are bimodally distributed.  The dynamics of this system has been the subject of longstanding interest.  In 1984 Kuramoto proposed several conjectures about its behavior; ten years later, Crawford obtained the first analytical results by means of a local center manifold calculation.  Nevertheless, many questions have remained open, especially about the possibility of global bifurcations.  Here we derive the system's  stability diagram for the special case where the bimodal distribution consists of two equally weighted Lorentzians.  Using an ansatz recently discovered by Ott and Antonsen, we show that in this case the infinite-dimensional problem reduces exactly to a flow in four dimensions.  Depending on the parameters and initial conditions, the long-term dynamics evolves to one of three states: incoherence, where all the oscillators are desynchronized; partial synchrony, where a macroscopic group of phase-locked oscillators coexists with a sea of desynchronized ones; and a standing wave state, where two counter-rotating groups of phase-locked oscillators emerge.  Analytical results are presented for the bifurcation boundaries between these states. Similar results are also obtained for the case in which the bimodal distribution is given by the sum of two Gaussians.
 
\end{abstract}
\maketitle
\section{Introduction}\label{section1}

\subsection{Background}\label{section1A}
Large systems consisting of many coupled oscillatory units occur
in a wide variety of situations \cite{Yamaguchi}. Thus the study of
the behaviors that such systems exhibit
has been an active and continuing area of research. An important
early contribution in this field was the introduction in 1975 by
Kuramoto \cite{Kuramoto75, Kuramoto84} of a simple model which illustrates
striking features of such systems. Kuramoto employed two
key simplifications in arriving at his model: (i) the coupling
between units was chosen to be homogeneous and all-to-all (i.e., `global'), so
that each oscillator would have an equal effect on all other oscillators;
and (ii) the oscillator states were solely described by a phase
angle $\theta (t)$, so that their uncoupled dynamics obeyed the
simple equation $d\theta _i/dt=\omega _i$, where $\omega _i $ is
the intrinsic natural frequency of oscillator $i$, $N\gg 1$ is the
number of oscillators, and $i=1,2,\ldots ,N$. The
natural frequencies $\omega _i$ are, in general, different for
each oscillator and are assumed to be drawn from some prescribed
distribution function $g(\omega )$.  

Much of the research on the Kuramoto model has focused on the case where  $g(\omega )$ is unimodal (for reviews of this literature, see \cite{Strogatz00, Ott02, Acebron05}).  Specifically, $g$ is usually assumed to be
symmetric about a maximum at frequency $\omega =\omega _0$ and to
decrease monotonically and continuously to zero as
$|\omega -\omega _0|$ increases. In that case, it
was found that as the coupling strength $K$
between the oscillators increases from zero in the large-$N$ limit, there
is a continuous transition at a critical coupling strength $K_c = 2/(\pi g(\omega_0))$.
For $K$ below $K_c$, the average macroscopic, time-asymptotic
behavior of the system is such that the oscillators in the system
behave incoherently with respect to each other, and an order
parameter (defined in Sec.~\ref{section2}) is correspondingly zero.
As $K$ increases past $K_c$,
the oscillators begin to influence each other in such a way that
there is collective global organization in the phases of the
oscillators, and the time-asymptotic order parameter assumes a
non-zero constant value that increases continuously for $K>K_c$ \cite{Kuramoto84, Strogatz00, Ott02, Acebron05, Strogatz92}.

It is
natural to ask how these results
change if other forms of $g(\omega )$ are considered. In this
paper we will address this question for what is perhaps the
simplest choice of a non-unimodal frequency distribution: we consider a distribution $g(\omega )$ that has two
peaks \cite{Barreto,Montbrio1}  and is the sum of two identical unimodal
distributions $\hat g$, such that $g(\omega
)=\frac{1}{2}[\hat g(\bar \omega -\omega _0)+\hat g(\bar \omega
+\omega _0)]$. We find that this modification to the original
problem introduces qualitatively new behaviors. As might be
expected, this problem has been previously
addressed \cite{Kuramoto84, Crawford}. However, due to its difficulty, the
problem was not fully solved, and, as we shall show, notable
features of the behavior were missed. 

\subsection{Reduction method}\label{section1B}

The development that makes
our analysis possible is the recent paper of Ott and
Antonsen \cite{Ott-Ant}. Using the method proposed in
Ref.~\cite{Ott-Ant} we reduce the original problem formulation
from an integro-partial-differential equation \cite{Strogatz00, Ott02, Strogatz92}
for the oscillator distribution function (a function of $\omega
,\theta $ and $t$) to a system of just a few ordinary differential
equations (ODEs). Furthermore, we analyze the reduced ODE system
to obtain its attractors and the bifurcations they experience with
variation of system parameters. 

The reduced ODE system, however,
represents a special restricted class of all the possible
solutions of the original full system \cite{Ott-Ant}. Thus a
concern is that the reduced system might miss some of the actual
system behavior. In order to check this, we have done numerical
solutions of the full system. The result is that, in all cases
tested, the time-asymptotic attracting behavior of the full system
and the observed attractor bifurcations are all contained in, and are
quantitatively described by, our ODE formulation.  
Indeed a similar result applies for the application of the method of Ref.~\cite{Ott-Ant} to the original Kuramoto model with unimodally distributed frequencies  \cite{Kuramoto75, Kuramoto84} and to the problem of the forced Kuramoto model with periodic drive \cite{Antonsen}.

On the other hand, the reduction method has not been mathematically proven to capture all the attractors, for any of the systems to which it has been applied~\cite{Ott-Ant, Antonsen}.  Throughout this paper we operate under the assumption (based on our numerical evidence) that the reduction method is reliable for the bimodal Kuramoto model.  But we caution the reader that in general the situation is likely to be subtle and system-dependent; see Sec.~\ref{section6D1} for further discussion of the scope and limits of the reduction method.  

\subsection{Outline of the paper}\label{section1C}

The organization of this paper is as follows. In Sec.~\ref{section2} we
formulate the problem and reduce it to the above-mentioned ODE
description for the case where $g(\omega)$ is a sum of Cauchy-Lorentz distributions.

Sec.~\ref{section3} provides an analysis of the ODE system. The main
results of Sec.~\ref{section3} are a delineation of the different types of
attractors that can exist, the regions of parameter space that
they occupy (including the possibility of bistability and
hysteresis), and the types of bifurcations that the attractors
undergo.

In Sec.~\ref{section4}, we establish that the attractors of the ODEs
obtained in Section \ref{section3} under certain symmetry assumptions
are attractors of the full ODE system. In Section \ref{section5},
we confirm that these attractors and bifurcations are also present
in the original system. In addition, we investigate the case where $g(\omega )$
is a sum of Gaussians, rather than Cauchy-Lorentz distributions. We find that the attractors
and bifurcations in the Lorentzian case and in the Gaussian case are of the
same types and that parameter space maps of the different behaviors are qualitatively
similar for the two distributions.

Finally, in Sec.~\ref{section6} we compare our results to the earlier work of Kuramoto \cite{Kuramoto84} and Crawford \cite{Crawford}.  Then we discuss the scope and limits of the reduction method used here, and offer suggestions for future research.

\section{Governing Equations}\label{section2}
\subsection{Problem definition}\label{section2A}
\begin{figure}[ht]
	\begin{center}
		\includegraphics[width=.5\textwidth]{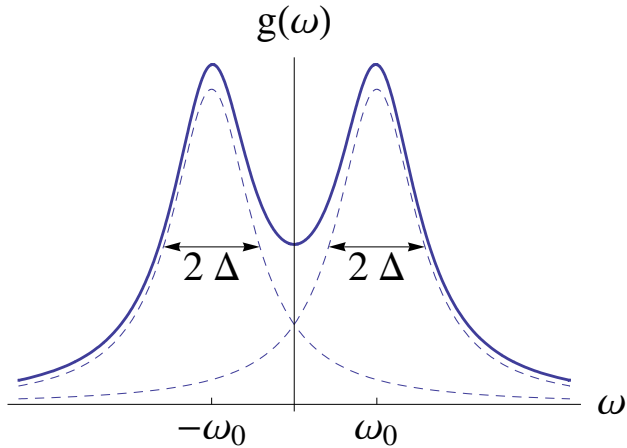}\
	\end{center}
\caption[Bimodal distribution of natural frequencies.]{A bimodal distribution of natural frequencies, $g(\omega)$, consisting of the sum of two Lorentzians.}
\label{fig:bimodaldist}
\end{figure}

We study the Kuramoto problem of $N$ oscillators with natural frequencies $\omega_i$,
\begin{eqnarray}\label{eq:goveqns}
 	\frac{d \theta_i (t)}{d t} &=& \omega_i + \frac{K}{N}\sum_{j=1}^N \sin{\left(\theta_j (t)-\theta_i (t)\right)},\
\end{eqnarray}
where $\theta_i$ are the phases of each individual oscillator and $K$ is the coupling strength. We study this system in the limit $N\rightarrow \infty$ for the case in which the distribution of natural frequencies is given by the sum of two Lorentzian distributions:\begin{equation}
 	g(\omega) = \frac{\Delta}{2\pi} \left(\frac{1}{(\omega-\omega_0)^2+\Delta^2}+\frac{1}{(\omega+\omega_0)^2+\Delta^2}\right).
\label{bimodaldist}
\end{equation}
Here $\Delta$ is the width parameter (half-width at half-maximum) of each Lorentzian and $\pm \omega_0$ are their center frequencies, as displayed in Fig.~\ref{fig:bimodaldist}.   A more physically relevant interpretation of $\omega_0$ is as the \emph{detuning} in the system (proportional to the separation between the two center frequencies).  
Note that we have written the distribution $g(\omega)$ so that it is symmetric about zero; this can be achieved without loss of generality by going into a suitable rotating frame.  

Another point to observe is that $g(\omega)$ is bimodal if and only if the peaks are sufficiently far apart compared to their widths.  Specifically, one needs $\omega_0 > \Delta/\sqrt{3}$.  Otherwise the distribution is unimodal and the classical results of  \cite{Kuramoto75, Kuramoto84,Strogatz00, Ott02} would still apply.  
   
\subsection{Derivation}\label{section2B}
In the limit where $N\rightarrow \infty$, Eq.~(\ref{eq:goveqns}) can be written in a continuous
formulation \cite{Kuramoto84, Strogatz00, Ott02} in terms of a probability density $f(\theta,\omega,t)$. Here $f$ is defined such that at time $t$, the fraction of oscillators with phases between $\theta$ and $\theta + d\theta$ and natural frequencies between $\omega$ and $\omega + d \omega$ is given by $f(\theta, \omega, t) \, d\theta \, d\omega$.  Thus 
\begin{equation}\label{normalization}
 \int_{-\infty}^\infty \int_0^{2 \pi} f(\theta, \omega,t)  \, d \theta \,d \omega  = 1
\end{equation}
and
\begin{equation}\label{freqdist}
\int_0^{2 \pi} f(\theta, \omega, t) \, d \theta = g(\omega),
\end{equation}
by definition of $g(\omega)$.  

The evolution of $f$ is given by the continuity equation 
describing the conservation of oscillators: 
\begin{eqnarray}\label{eq:continuityeqn}
 	\frac{\partial f}{\partial t} + \frac{\partial}{\partial \theta}\left(f v\right) &=& 0,\
\end{eqnarray}
where $v(\theta,\omega,t)$ is the angular velocity of the oscillators. From Eq.~(\ref{eq:goveqns}), we have
\begin{eqnarray}\label{eq:velocity}
 	v(\theta,\omega,t) &=& \omega + K \int_0^{2\pi} f(\theta',\omega,t) \sin(\theta'-\theta)  d\theta'.\
\end{eqnarray}
Following Kuramoto, we define a complex order parameter
\begin{eqnarray}\label{eq:complexOP}
	z(t)&=& \int_{-\infty}^{\infty} \int_0^{2\pi} e^{i\theta} f(\theta, \omega,t)  \, d \theta \,d \omega \
\end{eqnarray}
whose magnitude $|z(t)|\leq 1$ characterizes the degree to which the oscillators are bunched in phase,
and $\arg{(z)}$ describes the average phase angle of the oscillators. Expressing the velocity (\ref{eq:velocity}) in terms of $z$ we obtain
\begin{eqnarray}\label{eq:velocity2}
 	v(\theta,\omega,t) &=& \omega + K \,\textrm{Im}[z e^{-i\theta}] \\ &=& \omega + \frac{K}{2i} (z e^{-i\theta} - z^*e^{i\theta})\
\end{eqnarray}
where the * denotes complex conjugate.  

Following Ott and Antonsen \cite{Ott-Ant}, we now restrict attention to a special class of density functions. By substituting a Fourier series of the form 
\begin{equation}\label{fourier}
f(\theta, \omega, t) = \frac{g(\omega)}{ 2 \pi} \left[ 1 + \sum_{n=1}^\infty \left( f_n(\omega, t) e^{i n \theta} + \rm{c.c.} \right) \right],
\end{equation}
where `c.c' stands for the complex conjugate of the preceeding term,
and imposing the ansatz that 
\begin{equation}\label{eq:poisson}
f_n(\omega,t) = \alpha(\omega,t)^n,
\end{equation}
we obtain
\begin{eqnarray}\label{eq:amplitudeeqn}
 	\frac{\partial \alpha}{\partial t} + \frac{K}{2}(z\alpha^2-z^*)+i\omega\alpha &=& 0,
\end{eqnarray}
where
\begin{eqnarray}\label{eq:complexOP2}
	z^* &=& \int_{-\infty}^{\infty} \alpha(t,\omega)g(\omega)d\omega.\
\end{eqnarray}

We now consider solutions of (\ref{eq:amplitudeeqn}) and (\ref{eq:complexOP2}) for initial conditions $\alpha(\omega,0)$ that satisfy the following additional conditions: (i) $|\alpha(\omega,t)|\leq 1$; (ii) $\alpha(\omega,0)$ is analytically continuable into the lower half plane $\rm{Im}(\omega)<0$; and (iii) $|\alpha(\omega,t)|\rightarrow 0$ as $\rm{Im}(\omega)\rightarrow -\infty$. If these 
conditions are satisfied for $\alpha(\omega,0)$, then, as shown in \cite{Ott-Ant}, they continue to be satisfied by $\alpha(\omega,t)$ as it evolves under Eqs. (\ref{eq:amplitudeeqn}) and 
(\ref{eq:complexOP2}). Expanding $g(\omega)$ in partial fractions as
\begin{eqnarray}\nonumber
 g(\omega) &=& \frac{1}{4\pi i} \bigg[\frac{1}{(\omega-\omega_0)-i\Delta} - \frac{1}{(\omega-\omega_0)+i\Delta}+ \frac{1}{(\omega+\omega_0)-i\Delta} - \frac{1}{(\omega+\omega_0)+i\Delta} \bigg],\
\end{eqnarray}
we find it has four simple poles at $\omega = \pm \omega_0 \pm i\Delta$.
Evaluating (\ref{eq:complexOP2}) by deforming the integration path from the real $\omega$-axis to $\rm{Im}(\omega)\rightarrow -\infty$, the order parameter becomes
\begin{eqnarray}
 z(t) &=& \frac{1}{2}\left(z_1(t)+z_2(t)\right),\
\end{eqnarray}
where
\begin{eqnarray}
 z_{1,2}(t) &=& \alpha^*(\pm\omega_0-i\Delta,t). \
\end{eqnarray}

Substitution of this expression into (\ref{eq:amplitudeeqn}) yields two coupled complex ODEs, describing the evolution of two `sub'-order parameters,
\begin{eqnarray}\label{eq:fullgoveqn1}\nonumber
 \dot{z}_1 &=& - (\Delta + i \omega_0)z_1 \\
	   &&+ \frac{K}{4} \left[z_1 + z_2 -( z_1^* + z_2^*)z_1^2  \right]\\\nonumber &\\\label{eq:fullgoveqn2}\nonumber
 \dot{z}_2 &=& - (\Delta - i \omega_0)z_2 \\
	   && + \frac{K}{4} \left[z_1 + z_2 -(z_1^* + z_2^*)z_2^2 \right],\
\end{eqnarray}
where we use dots to represent the time derivative from now on. (This system agrees with
the results of \cite{Ott-Ant} for the case of two equal groups of oscillators with uniform coupling strength
and average frequencies $\omega_0$ and $-\omega_0$.)

\subsection{Reductions of the system}\label{section2C}
The system derived so far is four-dimensional. If we introduce polar coordinates $z_j=\rho_j e^{i\phi_j}$ and define the phase difference $\psi = \phi_2 - \phi_1$,
the dimensionality can be reduced to three:
\begin{eqnarray}
 \dot{\rho_1} &=& -\Delta \rho_1 + \frac{K}{4} \, (1-\rho_1^2)(\rho_1+\rho_2\cos{\psi})   \label{eq:goveqns3Da} \\
 \dot{\rho_2} &=& -\Delta \rho_2 + \frac{K}{4} \, (1-\rho_2^2)(\rho_1\cos{\psi}+\rho_2)   \label{eq:goveqns3Db} \\
 \dot{\psi} &=& 2\omega_0 - \frac{K}{4} \, \frac{\rho_1^2 + \rho_2^2 + 2 \rho_1^2\rho_2^2}{\rho_1\rho_2}\sin{\psi}.   \label{eq:goveqns3Dc} \
\end{eqnarray}

To facilitate our analysis we now look for solutions of Eqs.~(\ref{eq:goveqns3Da}-\ref{eq:goveqns3Dc}) that satisfy the symmetry condition
\begin{eqnarray}\label{eq:symmetry}
 \rho_1(t) =\rho_2(t) &\equiv& \rho(t).\
\end{eqnarray}
In Sec.~\ref{section4} we will verify that these symmetric solutions are stable to perturbations away from the symmetry manifold and that the attractors of Eqs.~(\ref{eq:fullgoveqn1}, \ref{eq:fullgoveqn2}) lie within this manifold.

Our analysis of the problem thus reduces to a study in the phase plane:
\begin{eqnarray}
	\dot{\rho} &=& \frac{K}{4}\rho\left(1-\frac{4\Delta}{K} - \rho^2  + (1-\rho^2)\cos{\psi}\right) \label{eq:goveqnprelima}\\
  	\dot{\psi} &=& 2\omega_0 - \frac{K}{2}\,(1+\rho^2)\sin{\psi}.\label{eq:goveqnprelimb}\
\end{eqnarray}

\section{Bifurcation Analysis}\label{section3}
\begin{figure}[ht]
	\begin{center}
		\includegraphics[width=\textwidth]{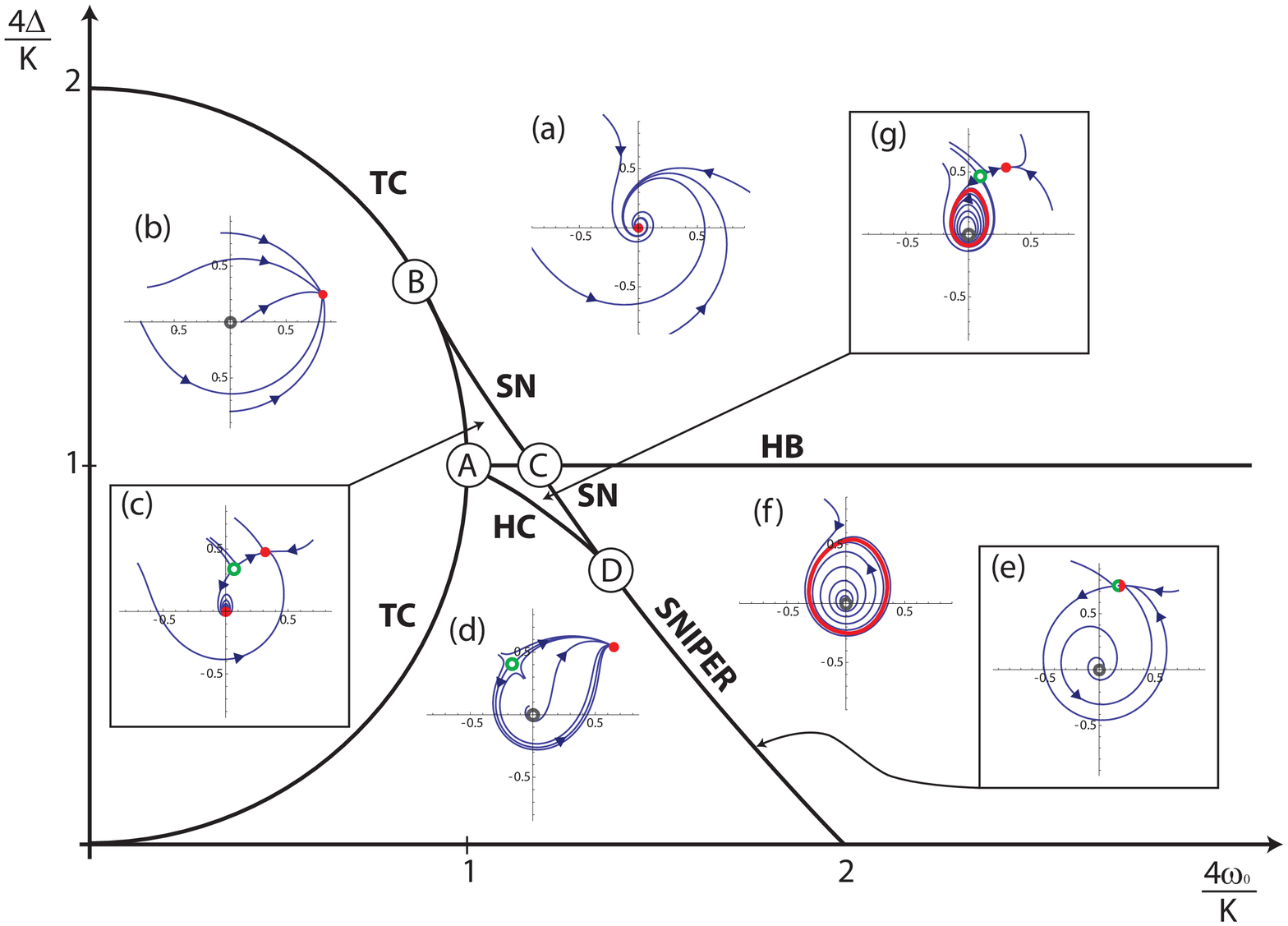}\
	\end{center}
\caption[What]{The bifurcation diagram for the Kuramoto system with a bimodal frequency distribution consisting of
two equally weighted Lorentzians. The various bifurcation curves are denoted as follows: TC=transcritical,
SN=saddle-node, HB=(degenerate) Hopf, HC=homoclinic, and SNIPER=Saddle-node-infinite-period. The insets,
labeled (a)-(g), show $(q,\psi)$ phase portraits in polar coordinates corresponding
to the regions where the insets are located (see arrows for the boxed insets).
Solid red dots and loops denote stable fixed point and limit cycles, respectively; open green dots are
saddle fixed points, and open gray circles are repelling fixed points. All parameters refer to their
original (unscaled) versions.
}
\label{fig:mainbifdiag1}
\end{figure}

Figure \ref{fig:mainbifdiag1} summarizes the results of our analysis of Eqs.~(\ref{eq:goveqnprelima}, \ref{eq:goveqnprelimb}).  We find that three types of attractors occur: the well-known incoherent and partially synchronized states \cite{Kuramoto75, Kuramoto84,Strogatz00, Ott02, Acebron05} corresponding to fixed points of (\ref{eq:goveqnprelima}, \ref{eq:goveqnprelimb}), as well as a standing wave state \cite{Crawford} corresponding to limit-cycle solutions.  In addition, we will show that the transitions between these states are mediated by transcritical, saddle-node, Hopf, and homoclinic bifurcations, as well as by three points of higher codimension.

\subsection{Scaling}\label{section3O}
To ease the notation we begin by scaling Eqs.~(\ref{eq:goveqnprelima}, \ref{eq:goveqnprelimb}).  
If we define $q=\rho^2$ and non-dimensionalize the parameters and time such that 
\begin{eqnarray}\label{eq:scaling}\nonumber
	\tilde t &=& \frac{K}{2}t\\
	\tilde\Delta &=& \frac{4\Delta}{K}\\\nonumber
	\tilde \omega_0&=&\frac{4\omega_0}{K}\
\end{eqnarray}
we obtain the dimensionless system
\begin{eqnarray}
	\dot{q} &=& q\left(1 -\Delta   -q + (1-q)\cos\psi\right) \label{eq:goveqns2Da}\\
  	\dot{\psi} &=& \omega_0 - (1+q)\sin{\psi}.\label{eq:goveqns2Db}\
\end{eqnarray}
Here the overdot now means differentiation with respect to dimensionless time, and we have dropped all the tildes for convenience. For the rest of this section, all parameters will be assumed to be dimensionless (so there are implicitly tildes over them) unless stated otherwise.

\subsection{Bifurcations of the incoherent state}\label{section3A}
The \emph{incoherent state} is defined by $\rho_1=\rho_2=0$, or by $q=0$ in the phase plane formulation. The linearization of the incoherent state, however, is
most easily performed in Cartesian coordinates using the formulation in Eqs.~(\ref{eq:fullgoveqn1}) and (\ref{eq:fullgoveqn2}). We find the degenerate eigenvalues
\begin{eqnarray}
 	\lambda_1 = \lambda_2 &=& 1-\Delta - \sqrt{1-\omega_0^2}\\
	\lambda_3 = \lambda_4 &=& 1-\Delta + \sqrt{1-\omega_0^2}.
\label{eq:originEVals}
\end{eqnarray}
This degeneracy is expected because the origin is \emph{always} a fixed point and because of the rotational invariance of that state.
It follows that the incoherent state is stable if and only if the real parts of the eigenvalues are less than or equal to zero. 

The boundary of stable incoherence therefore occurs when the following conditions are met:
\begin{equation*}\label{eq:StabCondOrigin}
 \left\{ \begin{array}{ll}
         \Delta = 1 +\sqrt{1-\omega_0^2} & \mbox{for $ \omega_0\leq 1$} \\
         \Delta = 1 & \mbox{for $\omega_0>1$}.\\
         \end{array} \right.\\
\end{equation*}
These equations define the semicircle and the half-line shown in Fig.~\ref{fig:mainbifdiag1}, labeled TC (for transcritical) and HB (for Hopf bifurcation), respectively.
(Independent confirmation of these results can be obtained from the continuous formulation of Eq.~(\ref{eq:goveqns}) directly, as shown in the Appendix.)
More precisely, we find that crossing the semicircle corresponds to a degenerate transcritical bifurcation, while
crossing the half-line corresponds to a degenerate supercritical Hopf bifurcation. 

In the latter case, the associated limit-cycle
oscillation indicates that the angle $\psi$ increases without bound; this reflects an increasing \emph{difference} between the phases of the two `sub'-order parameters of Eqs.~(\ref{eq:fullgoveqn1}, \ref{eq:fullgoveqn2}). In terms of the original model, this means that the oscillator population splits into two counter-rotating groups, each consisting of a macroscopic number of oscillators with natural frequencies close to one of the two peaks of $g(\omega)$.   Within each group the oscillators are frequency-locked. Outside the groups the oscillators remain desynchronized, drifting relative to one another and to the locked groups.   This is the state Crawford \cite{Crawford} called a \emph{standing wave}.  Intuitively speaking, it occurs when the two humps in the frequency distribution are sufficiently far apart relative to their widths. In Kuramoto's vivid terminology \cite{Kuramoto84}, the population has spontaneously condensed into ``a coupled pair of giant oscillators.''

\subsection{Fixed point solutions and saddle-node bifurcations}\label{section3B}
\begin{figure}[ht]
	\begin{center}
		\includegraphics[width=.7\textwidth]{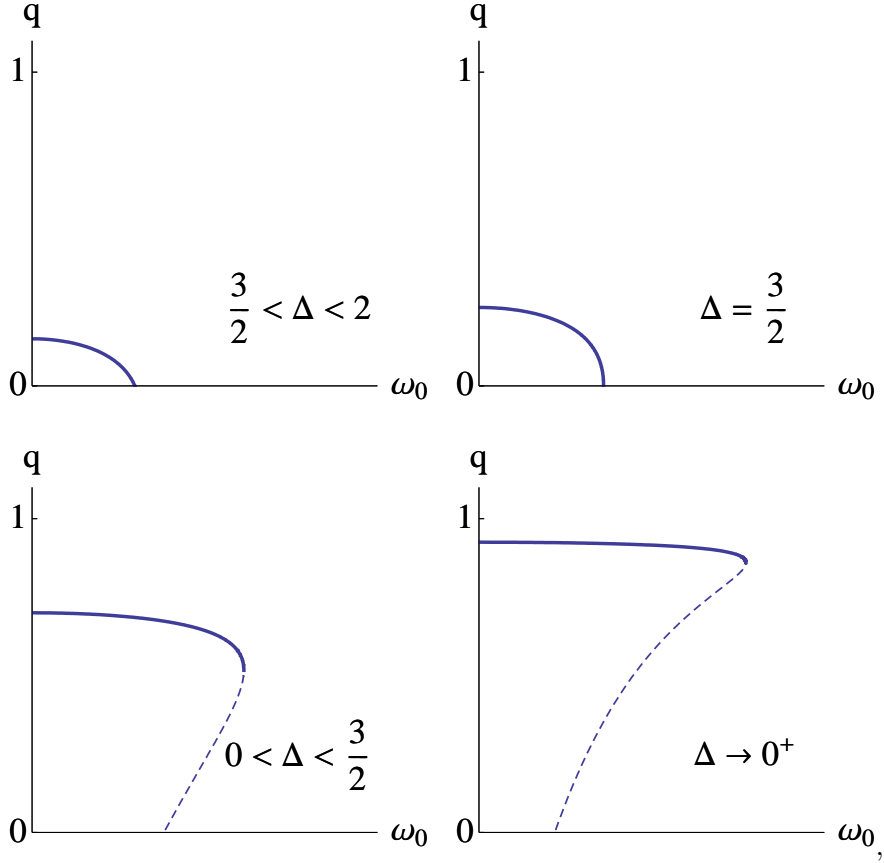},\
	\end{center}
\caption[Fold bifurcation.]{Saddle-node bifurcation: at $\Delta = 3/2$, $q$ becomes double-valued.}
\label{fig:foldbif}
\end{figure}

Along with the trivial incoherent state $q=0$, the other fixed points of Eqs.~(\ref{eq:goveqns2Da}, \ref{eq:goveqns2Db}) satisfy $1-\Delta-q=(q-1)\cos\psi$, and $\omega_0 = (q+1)\sin\psi$. Using trigonometric identities, we obtain
\begin{eqnarray}
 	1 &=& \left(\frac{\omega_0}{q+1}\right)^2 + \left(\frac{1-\Delta-q}{q-1}\right)^2,\
\end{eqnarray}
or equivalently,
\begin{eqnarray}\label{eq:fpsurfomega}
 	\omega_0 &=&\pm \frac{1+q}{1-q}\sqrt{\Delta(2-2q-\Delta)}.\
\end{eqnarray}
Thus, the fixed point surface $q=q(\omega_0,\Delta)$ is defined implicitly. It can be single- or double-valued as a function of $\omega_0$ for fixed $\Delta$. To see this, consider how $\omega_0$ behaves as $q\rightarrow 0^+$. We find that 
\begin{eqnarray}
 	\omega_0 &\sim& \sqrt{\Delta(2-\Delta)} \left[ 1+\frac{3-2\Delta}{2-\Delta}q+\mathcal{O}(q^2) \right], 
\end{eqnarray}
from which we observe that the behavior changes qualitatively at $\Delta=3/2$, as shown in Fig.~\ref{fig:foldbif}.
\begin{figure}[ht]
	\begin{center}
 		\includegraphics[width=0.35\textheight]{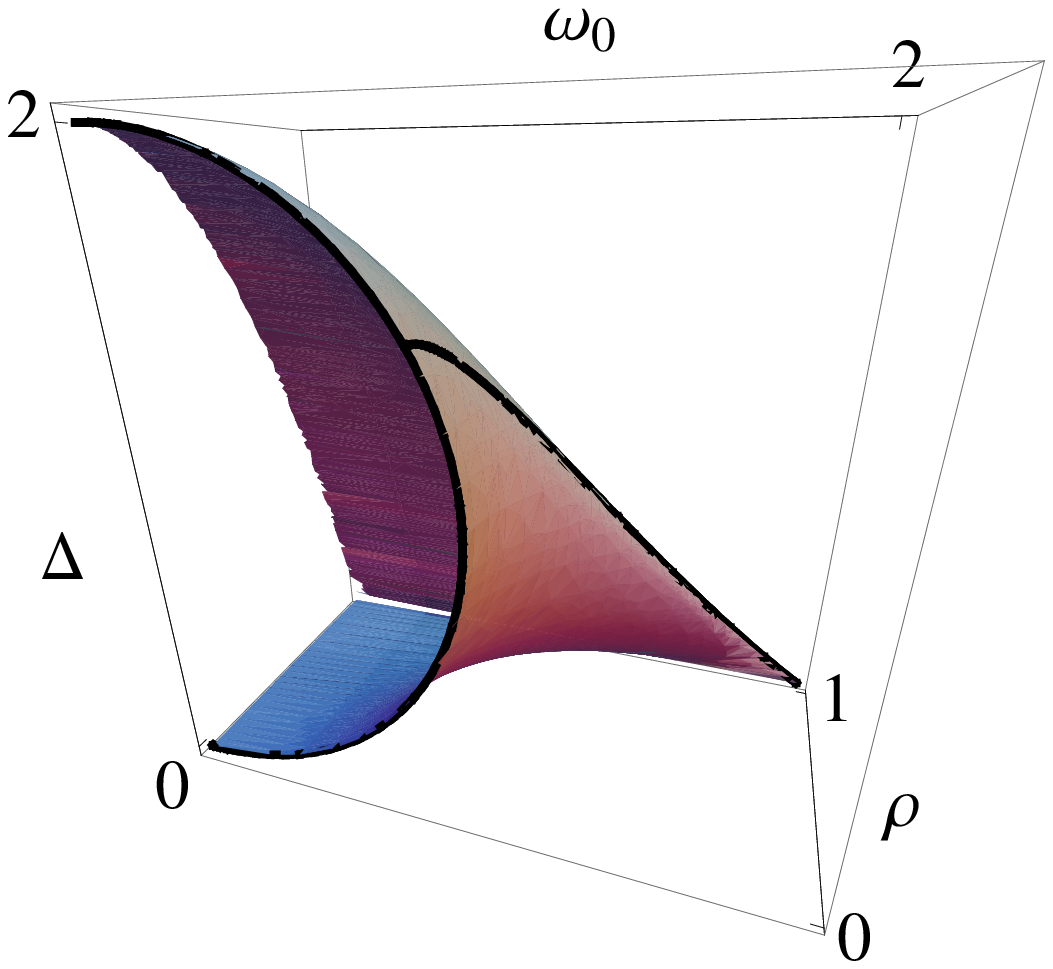}\
	\end{center}
\caption[Fixed point surface.]{Fixed point surface. Bifurcation curves at the origin and the saddle-node curve are emphasized in black.}
\label{fig:ParsolGequal}
\end{figure}

The surface defined by $\rho=\rho(\omega_0,\Delta)$ can be plotted parametrically using $\rho$ and $\Delta$, as is seen in Fig.~\ref{fig:ParsolGequal}.   The fold in the surface corresponds to a saddle-node bifurcation. Plots of the phase portrait of $(q,\psi)$ reveal that the upper branch of the double-valued surface in Fig.~\ref{fig:foldbif} corresponds to sinks, and the lower branch to saddle points; see Fig.~\ref{fig:mainbifdiag1} (c), (d), and (g).

In physical terms, the sink represents a stable \emph{partially synchronized state}, which is familiar from the classic Kuramoto model with a unimodal distribution \cite{Kuramoto84, Strogatz00, Ott02, Acebron05}.  The oscillators whose natural frequencies are closest to the center of the frequency distribution $g(\omega)$ become rigidly locked, and maintain constant phase relationships among themselves---in this sense, they act collectively like a ``single giant oscillator,'' as Kuramoto \cite{Kuramoto84} put it.  Meanwhile the oscillators in the tails of the distribution drift relative to the locked group, which is why one describes the synchronization as being only partial.   

The saddle points also represent partially synchronized states, though of course they are unstable.  Nevertheless they play an important role in the dynamics because they can annihilate the stable partially synchronized states; this happens in a saddle-node bifurcation along the fold mentioned above.  To calculate its location analytically, we use (\ref{eq:fpsurfomega}) and impose the condition for a turning point, $\partial\omega_0/\partial q = 0$, which yields
\begin{eqnarray}
 	q^2 - 4q + 3 -2\Delta &=&0.\
\end{eqnarray}
Eliminating $q$ from this equation using (\ref{eq:fpsurfomega}), we obtain the equation for the saddle-node bifurcation curve
\begin{eqnarray}\label{eq:SNcurve}
 	\omega_0 &=& \sqrt{2-10\Delta-\Delta^2+2(1+2\Delta)^{3/2}}.\
\end{eqnarray}
This curve is labeled SN in Fig.~\ref{fig:mainbifdiag1}. Its intersection with the semicircle TC occurs at $(\omega_0,\Delta)=(\frac{\sqrt{3}}{2},\frac{3}{2})$, and is labeled B in the figure. Note also that point C in the figure
is \emph{not} a Takens-Bagdanov point, as the saddle-node and Hopf bifurcations occur at different locations
in the state space; see Figs.~\ref{fig:mainbifdiag1} (a) and (g). 

\subsection{Bistability, homoclinic bifurcations, and SNIPER}\label{section3D}
An examination of the dynamics corresponding to the approximately triangular parameter
space region ABC in Fig.~\ref{fig:mainbifdiag1} shows bistability. More specifically,
we find that the stable incoherent fixed point coexists with the stable partially synchronized state
produced by the saddle-node bifurcation described above, as shown in the
state-space plot in Fig.~\ref{fig:mainbifdiag1}(c).  

Further study of these state-space plots
led us to the homoclinic
bifurcation curve marked HC, which was obtained numerically. The coexistence of states
continues into region ACD, where we found that the stable partially synchronized state now 
coexists with the stable limit cycle created at the Hopf curve. (See Fig.~\ref{fig:mainbifdiag1}(g).) This
limit cycle is then destroyed by crossing the homoclinic curve, which is
bounded by point A on one side and by point D on the other.

At point D, the homoclinic curve merges with the saddle-node curve. This codimension-two
bifurcation, occurring at approximately (1.3589, 0.7483), is known as a saddle-node-loop \cite{Guckenheimer}. Below D, however, the saddle-node curve exhibits an interesting feature: the saddle-node bifurcation
occurs on an invariant closed curve. This bifurcation scenario is known as a saddle-node infinite-period bifurcation, or in short, \emph{SNIPER}. If we traverse the SNIPER curve from left to right, the sink and saddle (the stable and unstable partially synchronized states) coalesce, creating a loop with infinite period. Beyond that, a stable limit cycle then appears---see Figs.~\ref{fig:mainbifdiag1} (d), (e), and (f).

In conclusion, we have identified six distinct regions in parameter space
and have identified the bifurcations that occur at the boundaries.

\section{Transverse Stability}\label{section4}
Our analysis so far has been based on several simplifying assumptions.  First, we restricted attention to a special family of oscillator distribution functions $f(\theta,\omega,t)$ and a bimodal Lorentzian form for $g(\omega)$, which  enabled us to reduce the original infinite-dimensional system to a three-dimensional system of ODEs, Eqs.~(\ref{eq:goveqns3Da}-\ref{eq:goveqns3Dc}). Second, we considered only symmetric solutions of these ODEs, by assuming $\rho_1 = \rho_2$; this further decreased the dimensionality from three to two.  

The next two sections test the validity of these assumptions. We begin here by showing that the non-zero fixed point attractor (the stable partially synchronized state) and the limit cycle attractor (the standing wave state) for Eqns.~(\ref{eq:goveqns2Da}, \ref{eq:goveqns2Db}) are transversely stable to small symmetry-breaking perturbations, i.e., perturbations off the invariant manifold defined by  $\rho_1 = \rho_2$. This does not rule out the possible existence of attractors off this manifold, but it does mean that the attractors in the two-dimensional symmetric manifold are guaranteed to constitute attractors in the three-dimensional ODE system~(\ref{eq:goveqns3Da}-\ref{eq:goveqns3Dc}).  

Let $\kappa = K/4$ and consider the reduced governing equations (\ref{eq:goveqns3Da}-\ref{eq:goveqns3Dc}) without symmetry. Introducing the longitudinal and transversal variables 
\begin{eqnarray}\nonumber\label{eq:longtransdefns}
 	\rho_{\parallel}&=& \frac{1}{2}(\rho_1+\rho_2)\\
	\rho_{\bot}&=& \frac{1}{2}(\rho_1-\rho_2),\
\end{eqnarray}
and substituting these into (\ref{eq:goveqns3Da}-\ref{eq:goveqns3Dc}), we derive the equation for the transversal component
\begin{eqnarray}\label{eq:transversalevolution}\nonumber
 	\dot{\rho}_{\bot}&=& \rho_{\bot}\Big[(\kappa-\Delta) - \kappa(3\rho_{\parallel}^2+\rho_{\bot}^2)- \kappa\cos{\psi}(1+\rho_{\parallel}^2-\rho_{\bot}^2)\Big],\
\end{eqnarray}
which describes the order parameter dynamics off the symmetric manifold. 

To simplify the notation, let $q_{\parallel}=\rho_{\parallel}^2$ and $q_{\bot}=\rho_{\bot}^2$ and scale the system using Eqs.~(\ref{eq:scaling}), as before.  
Linearization
and evaluation at the asymptotic solution denoted by $(q_0,\psi_0)$, which may be either a fixed point or a limit cycle, yields the variational equation
\begin{eqnarray}\label{eq:transevaleqn}
 	\delta \dot{q}_{\bot} &=& \lambda_{\bot}\delta q_{\bot}\
\end{eqnarray}
where 
\begin{eqnarray}\label{eq:transversaleval}
 	\lambda_{\bot} &=& 1-\Delta - 3 q_0 - (1+q_0) \cos{\psi_0} .\
\end{eqnarray}
Observe that $\delta q_{\parallel}$ and $\delta\psi$ do not appear in linear order on the right hand side of (\ref{eq:transevaleqn}). This decoupling implies that $\lambda_{\bot}$ is the eigenvalue associated with the transverse perturbation $\delta q_{\bot}$, in the case where $q_0$ is a fixed point.  Similarly, if $q_0$ is a limit cycle, the Floquet exponent associated with $\delta q_{\bot}$ is simply $\langle\lambda_{\bot} \rangle$, where the brackets denote a time average over one period.
Hence the fixed point will be transversely stable if $\lambda_{\bot}<0$.  The analogous condition for the limit cycle is $\langle\lambda_{\bot} \rangle<0$.

\subsection{Fixed point stability}
To test the transverse stability of sinks for the two-dimensional flow, we solve  
Eq.~(\ref{eq:goveqns2Da}) for fixed points and obtain
\begin{eqnarray}
 	0&=& 1-\Delta - q_0 + (1-q_0)\cos{\psi_0}.\
\end{eqnarray}
Subtracting this from (\ref{eq:transversaleval}), we find
\begin{eqnarray}
 	\lambda_{\bot} &=& -2 (q_0 + \cos{\psi_0}).\
\end{eqnarray}
Hence $\cos{\psi_0}>0$ is a sufficient condition for transverse stability. But at a non-trivial fixed point, 
\begin{eqnarray}
 	\cos{\psi_0} &=& \frac{1-(\Delta+q_0)}{q_0-1},\
\end{eqnarray}
so the transverse stability condition is equivalent to $q_0+\Delta>1$. 

We claim that this inequality holds everywhere on the upper branch of the fixed point surface (\ref{eq:fpsurfomega}).
Obviously the inequality is satisfied at all points where $\Delta>1$. For all other cases, consider the turning point from Fig.~\ref{fig:foldbif} defined by $q_{sn}=2\pm\sqrt{1+2\Delta}$. Since the function of interest, $Q(\Delta)\equiv q_{sn}+\Delta$, has a global minimum with $Q(0)=1$, and $q_{sn}$ is independent of $\omega_0$ (at fixed $\Delta$), it is a lower bound for all $q(\omega_0)$ on the upper sheet of the fixed point surface, provided that $q(\omega_0)$ is monotonically decreasing on the interval of $[0,\omega_{sn}]$. In fact, it is easier to establish that $0>\partial\omega_0/\partial q = \Delta/D (q^2-4q+3-2\Delta)$ with $D=(q-1)^2\sqrt{2\Delta-2q\Delta-\Delta^2}$; the latter expression is positive, and $q^2-4q+3-2\Delta < 0$ whenever $1>q>q_{sn}$. Thus transverse stability for the nodes on the fixed point surface follows.

\subsection{Limit cycle stability}
To examine the transverse linear stability  of the limit cycle, we calculate the transverse Floquet exponent by averaging the eigenvalue over the period of one oscillation:
\begin{eqnarray}
 \langle \lambda_{\bot}\rangle &=& 1-\Delta - 3 \langle q_0 \rangle - \left(\langle\cos{\psi_0}\rangle+\langle q_0 \cos{\psi_0}\rangle \right).
\end{eqnarray}
In order to render this expression definite, we rewrite Eq.~(\ref{eq:goveqns2Da}) in terms of the limit cycle solution $(q_0,\psi_0)$:
\begin{eqnarray}
 	\frac{d}{dt}\left(\ln{q_0}\right) &=& 1-\Delta-q_0
+ (1-q_0)\cos{\psi_0}.\
\end{eqnarray}
Periodicity on the limit cycle guarantees $\langle \frac{d}{dt}\ln{q_0}\rangle = 0$, and so we have
\begin{eqnarray}
 	0 &=& 1-\Delta- \langle q_0 \rangle
+ \langle (1-q_0)\cos{\psi_0} \rangle,\
\end{eqnarray}
which we subtract from the averaged eigenvalue to yield
\begin{eqnarray}\label{eq:LCstability}
  \langle \lambda_{\bot}\rangle &=& -2 (\langle q_0 \rangle + \langle \cos{\psi_0} \rangle).\
\end{eqnarray}
Although we are not able to analytically demonstrate that $\langle \lambda_{\bot}\rangle$ in (\ref{eq:LCstability}) is negative, we have calculated  $\langle q_0\rangle$ and  $\langle \cos{\psi_0}\rangle$ numerically for the limit cycle attractors of Eqs.~(\ref{eq:goveqns3Da}-\ref{eq:goveqns3Dc}). This was done
for $2500$ parameter values corresponding to a grid in dimensionless parameter space, by sampling 50 evenly spaced values $\omega \in [0.01,2.5]$ and $\Delta \in [0.01,2.1]$.
The simulations were run with $N = 1024$ oscillators.  In all the cases that we tested, we found that $\langle \lambda_{\bot}\rangle<0$.

\section{Numerical Experiments}\label{section5}

All of the results described above were obtained using the reduced ODE models derived  in Sec.~\ref{section2} B and C, and are therefore subject to the
restrictions described therein. It is therefore reasonable to ask if these
results agree with the dynamics of the original system given in Eq.~(\ref{eq:goveqns}).  To check this, a series of direct simulations of Eq.~(\ref{eq:goveqns}) using $N = 10,000$ oscillators and fourth-order Runge-Kutta numerical integration were performed.

\begin{figure}[!ht]
	\begin{center}
		\includegraphics[width=0.7\textwidth]{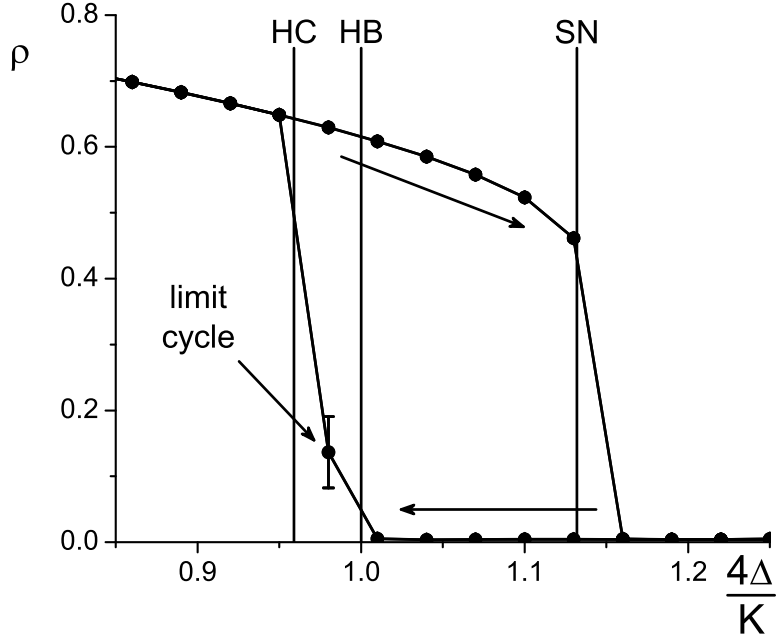}\
	\end{center}
\caption[]{Hysteresis loop as observed when traversing the bistable regions shown in Fig.~\ref{fig:mainbifdiag1}
in the directions shown (arrows) along the line at $4\omega_0/K=1.092$. The data were obtained from a simulation of Equation (\ref{eq:goveqns})
with $N=10,000$ and $K=1$. Vertical lines indicate where the reduced ODE models of Section \ref{section2} predict
homoclinic (HC), degenerate Hopf (HB), and saddle-node (SN) bifurcations. Note that the point marked
`limit cycle' has a large error bar, reflecting the oscillations in the order parameter.}
\label{fig:hysteresis}
\end{figure}

First, we compared solutions of Eq.~(\ref{eq:goveqns}) with those of our reduced system Eqs.~(\ref{eq:goveqnprelima}, \ref{eq:goveqnprelimb}) in the region where we predicted the coexistence of attractors. For example,  we show in Fig.~\ref{fig:hysteresis} a bifurcation diagram computed along the line $4\omega_0/K=1.092$ that traverses the region ABCD in Fig.~\ref{fig:mainbifdiag1}. (Note that here and for the rest of the paper, we revert to using the original, dimensional form of the variables.) The vertical lines in Fig.~\ref{fig:hysteresis} indicate the locations of the bifurcations that were identified using the ODE models. For each point plotted, the simulation was run until the order parameter exhibited its time-asymptotic behavior; this was then averaged over the subsequent 5000 time steps. Error bars denote standard deviation. Note in particular the hysteresis, as well as the point with the large error bar, indicating the predicted limit cycle behavior.

Next, we examined the behavior of Eq.~(\ref{eq:goveqns}) at 121 parameter values corresponding to an $11 \times 11$ regular grid superimposed on Fig.~\ref{fig:mainbifdiag1}, ranging from 0.1 to 2.1 at intervals of 0.2 on each axis. (In all cases, $K$ was set to 1, and $\Delta$ and $\omega_0$ were varied.) An additional series was run using a smaller grid (from 0.6 to 1.6 at intervals of 0.1 on each axis), to focus on the vicinity of region ABCD in Fig.~\ref{fig:mainbifdiag1}. Initial conditions were chosen systematically in 13 different ways, as follows:

\begin{enumerate}
\item The oscillator phases were uniformly distributed around the circle, so that the overall order parameter had magnitude $r=0$.

\item The oscillators were all placed in phase at the same randomly chosen angle in $[0, 2\pi]$, so that $r=1$.

\item The remaining 11 initial conditions were chosen by regarding the system as composed of two sub-populations, one for each Lorentzian in the bimodal distribution of frequencies, as in \cite{Barreto}.  In one of the sub-populations, the initial phases of the oscillators were chosen to be randomly spaced within the angular sector $[c+d, c-d]$, where $c$ was chosen randomly in $[0, 2\pi]$ and $d$ was chosen at random such that the sub-order parameter magnitude $r_1$ = 0.1, 0.2, 0.3, 0.4, 0.5, 0.6, 0.7, 0.8, or 0.9 (all approximately). The result was that $r_1$ had one of these magnitudes and its phase was random in $[0, 2\pi]$.  The same procedure was followed for the other sub-population, subject to the constraint that $r_1 \neq r_2$.  Our idea here was to deliberately break the symmetry of the system initially, to test whether it would be attracted back to the symmetric subspace defined by Eq.~(\ref{eq:symmetry}).

\end{enumerate}

\begin{figure}[!ht]
	\begin{center}
		\includegraphics[width=\textwidth]{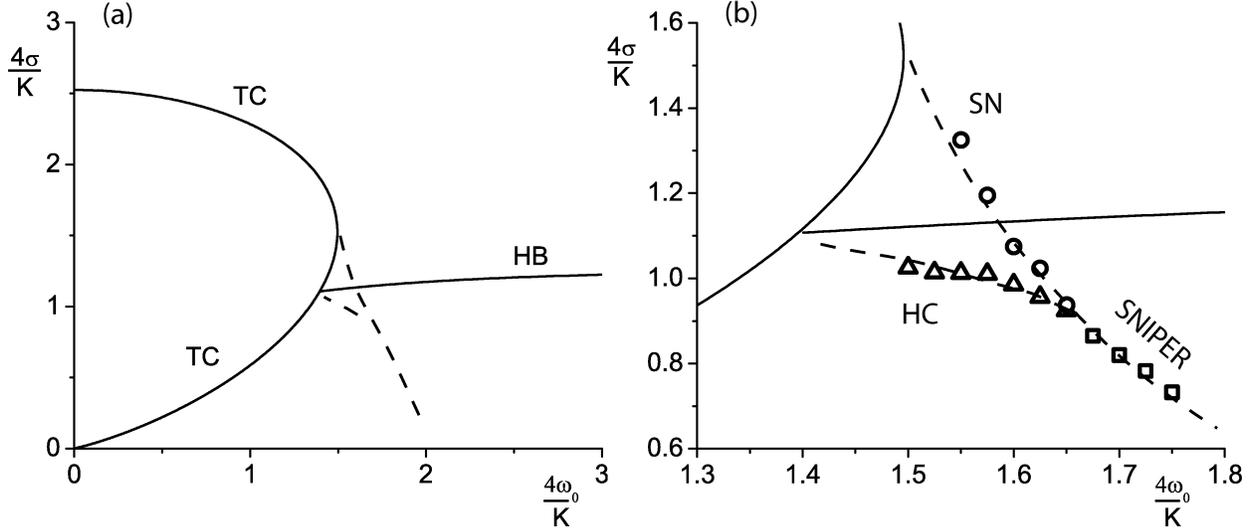}\
	\end{center}
\caption[]{(a) The bifurcation diagram for the Kuramoto system with a bimodal frequency distribution consisting of
two equally weighted Gaussians. All the features in Fig.~\ref{fig:mainbifdiag1} are present, but are somewhat
distorted. The transcritical (TC) and (degenerate) Hopf curves (HB) were obtained
as described in the Appendix. The dotted lines represent conjectured saddle-node, homoclinic, and SNIPER curves.
These are based on the numerically-observed bifurcations shown in (b), which is a magnification of
the central region of (a). The symbols represent saddle-node (circles), homoclinic (triangles),
and SNIPER (squares) bifurcations.} 
\label{fig:Gaussiandiag}
\end{figure}

In all the cases we examined, no discrepancies were found between the simulations and the predicted behavior. Although these tests were not exhaustive, and certainly do not constitute a mathematical proof, they are consistent with the conjecture that no additional attractors beyond those described in Section III exist.

We then  investigated the generality of our results by replacing the bimodal Lorentzian natural frequency
distribution, Eq.~(\ref{bimodaldist}), with the sum of two Gaussians:
\begin{equation}
g(\omega)=\frac{1}{\sigma \sqrt{2\pi}} \left( e^{-\frac{(\omega - \omega_0)^2}{2 \sigma^2}} + e^{-\frac{(\omega + \omega_0)^2}{2 \sigma^2}} \right)
\end{equation}
and computing  the corresponding bifurcation diagram analogous to Fig.~\ref{fig:mainbifdiag1}.  The results are shown in Fig.~\ref{fig:Gaussiandiag}. The transcritical (TC) and degenerate Hopf bifurcation (HB) curves were obtained using the continuous formulation of Eq.~(\ref{eq:goveqns});
see the Appendix for details. In addition, saddle-node, homoclinic, and SNIPER bifurcations
were numerically observed at several parameter values, and based on these data, we estimated
the location of the corresponding curves (dashed lines). All the features of Fig.~\ref{fig:mainbifdiag1}
are preserved, but the curves are somewhat distorted.

\section{Discussion}\label{section6}

We conclude by relating our work to three previous studies, and then offer suggestions for further research,  both theoretical and experimental.  

\subsection{Kuramoto's conjectures}
In his book on coupled oscillators, Kuramoto \cite{Kuramoto84} speculated about how the transition from incoherence to mutual synchronization might be modified if the oscillators' natural frequencies were bimodally distributed across the population.  On pp.75--76 of Ref. \cite{Kuramoto84}, he wrote ``So far, the nucleation has been supposed to be initiated at the center of symmetry of $g$.  This does not seem to be true, however, when $g$ is concave there.''  His reasoning was that for a bimodal system, synchrony would be more likely to start at the peaks of $g$.  If that were true, it would mean that a system with two equal peaks would go directly from incoherence to having two synchronized clusters of oscillators, or what we have called the standing wave state, as the coupling $K$ is increased.  The critical coupling at which this transition would occur, he argued, should be $K_c = 2/(\pi g(\omega_{\rm max}))$, analogous to his earlier result for the unimodal case.  According to this scenario, the synchronized clusters would be tiny at onset, comprised only of oscillators with natural frequencies near the peaks of $g(\omega)$.  Because of their small size, Kuramoto claimed these clusters ``will behave almost independently of each other.''  With further increases in $K$, however, the clusters ``will come to behave like a coupled pair of giant oscillators, and for even stronger coupling they will eventually be entrained to each other to form a single giant oscillator.''  (This is what we have called the partially synchronized state.)

Let us now re-examine Kuramoto's conjectures in light of our analytical and numerical results, as summarized in Fig.~\ref{fig:BD_overview}(a).  For a fair comparison, we must  assume that $g$ is concave at its center frequency $\omega =0$; for the bimodal Lorentzian (Eq. (\ref{bimodaldist}), this is equivalent to $\omega_0/\Delta > 1/\sqrt 3$.  (Otherwise $g$ is unimodal and incoherence bifurcates to partial synchronization as $K$ is increased, consistent with Kuramoto's classic result as well as the lowest portion of Fig.~\ref{fig:BD_overview}(a).)   

So restricting attention from now on to the upper part of Fig.\ref{fig:BD_overview}(a) where $\omega_0/\Delta > 1/\sqrt 3$, what actually happens as $K$ increases?  Was Kuramoto right that the bifurcation sequence is always incoherence $\rightarrow$ standing wave $\rightarrow$ partial synchronization?  

No.  For $\omega_0/\Delta$ between $1/\sqrt 3$ and $1$ (meaning the distribution is just barely bimodal), incoherence bifurcates directly to partial synchronization---the ``single giant oscillator'' state---without ever passing through an intermediate standing wave state.  In effect, the system still behaves as if it were unimodal.  But there is one new wrinkle:  we now see hysteresis in the transition between incoherence and partial synchronization, as reflected by the lower bistable region in Fig.~\ref{fig:BD_overview}(a).   

Is there any part of Fig.~\ref{fig:BD_overview}(a) where Kuramoto's scenario really does occur?  Yes---but it requires that the peaks of $g$ be sufficiently well separated.  Specifically, suppose $\omega_0/\Delta > 1.81\ldots$,  the value at the codimension-2 saddle-node-loop point where the homoclinic and SNIPER curves meet
(i.e., point D in Fig.~\ref{fig:mainbifdiag1}).  In this regime everything behaves as Kuramoto predicted.   

An additional subtlety occurs in the intermediate regime where the peaks of $g$ are neither too far apart nor too close together.  Suppose that $1< \omega_0/\Delta  < 1.81\ldots$.  Here the system shows a different form of hysteresis.  The bifurcations occur in the sequence that Kuramoto guessed as $K$ increases, but \emph{not} on the return path.  Instead, the system skips the standing wave state and dissolves directly from partial synchronization to incoherence as $K$ is decreased.  

Finally we note that Kuramoto's conjectured formula $K_c = 2/(\pi g(\omega_{\rm max}))$ is incorrect, although it becomes asymptotically valid in the limit of widely separated peaks.  Specifically, his prediction is equivalent to $K_c = \frac{8 \Delta}{1+\sqrt{1+(\Delta/\omega_0)^2}} \sim 4 \Delta (1-\frac{1}{4} (\Delta/\omega_0)^2)$, which approaches the correct result $K_c = 4 \Delta$ as $\omega_0/\Delta \rightarrow \infty$.

\subsection{Crawford's center manifold analysis}

Crawford \cite{Crawford} obtained the first mathematical results for the system studied in this paper.  Using center manifold theory, he calculated the weakly nonlinear behavior of the infinite-dimensional system in the neighborhood of the incoherent state.   From this he derived the stability boundary of incoherence.  His analysis also included the effects of white noise in the governing equations.  
\begin{figure}[!bh]
	\begin{center}
		\begin{tabular}{cc}
			\includegraphics[width=1\textwidth]{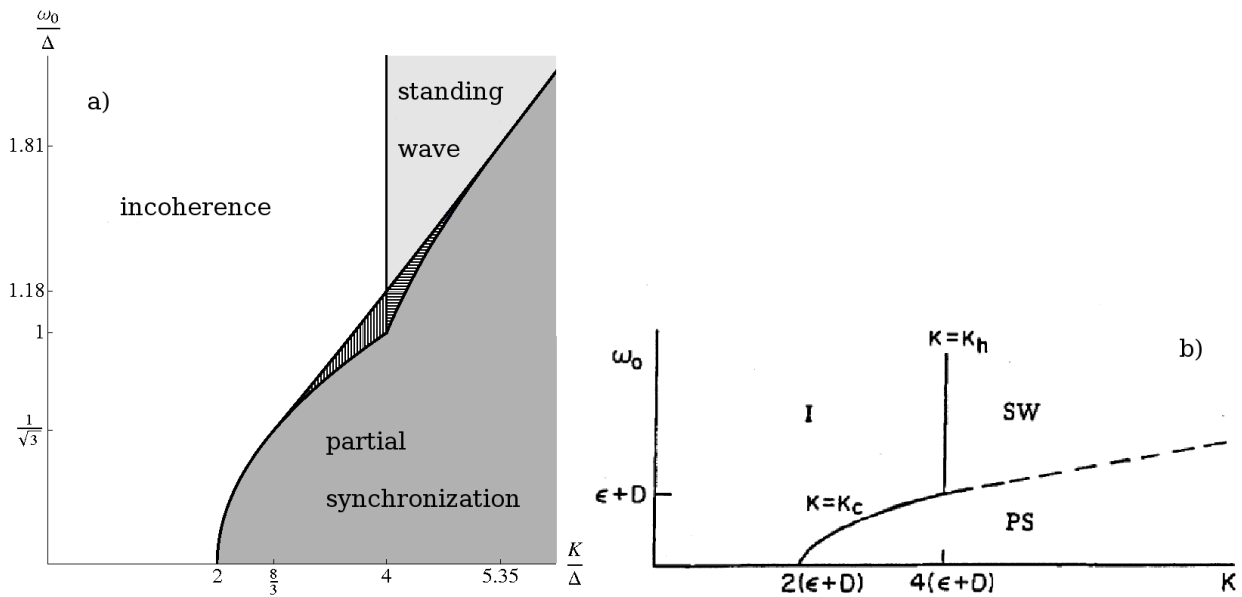}
		\end{tabular}
	\end{center}
\caption[Crawford's bifurcation diagram.]{Left: Results from our analysis. \emph{white}: incoherence, \emph{dark gray}: partial synchronization, \emph{light gray}: standing wave (limit cycles), \emph{vertical lines}: coexistence of incoherent and partially synchronized states, \emph{horizontal lines}: coexistence of partial synchronization and standing waves.
Right: Crawford's bifurcation diagram in \cite{Crawford}. In our study there is no noise, and so the diffusion is $D=0$. Crawford's $\epsilon$ corresponds to our $\Delta$. \emph{I}: Incoherent states, \emph{PS}: partially synchronized, \emph{SW}: standing wave, equivalent to what we describe as two counterrotating flocks of oscillators. (Permission to print by Springer Verlag.)
}
\label{fig:BD_overview}
\end{figure}

Figure~\ref{fig:BD_overview}(b), reproduced from Fig.~4 in Ref.~\cite{Crawford}, summarizes Crawford's findings.  Here $D$ is the noise strength (note: our analysis is limited to $D=0$), $\epsilon$ is the width of the Lorentzians (equivalent to $\Delta$ in our notation), and $\pm \omega_0$ are the center frequencies of the Lorentzians (as here).  The dashed line in Fig.~\ref{fig:BD_overview}(b) shows Crawford's schematic depiction of the unknown stability boundary between the standing waves and the partially synchronized state.  He suggested a strategy for calculating this boundary, and highlighted it as an open problem, writing in the figure caption, ``...the precise nature and location of this boundary have not been determined."  Our results, summarized in Figs.~\ref{fig:mainbifdiag1} and Fig.~\ref{fig:BD_overview}(a), now fill in the parts that were missing from Crawford's analysis.

\subsection{Stochastic model of Bonilla et al.}

In a series of papers (see \cite{Acebron05} for a review), Bonilla and his colleagues have explored what happens if one replaces the Lorentzians in the frequency distribution with $\delta$-functions, and adds white noise to the governing equations.  The resulting system can be viewed as a stochastic counterpart of the model studied here; in effect, the noise blurs the $\delta$-functions into bell-shaped distributions analogous to Lorentzians or Guassians.  And indeed, the system shows much of the same phenomenology as seen here: incoherence, partially synchronized states, standing waves, and bistability  \cite{Acebron05}.  

However, a complete bifurcation diagram analogous to Fig.~\ref{fig:mainbifdiag1} has not yet been worked out for this model.  The difficulty is that no counterpart of the ansatz  (\ref{eq:poisson})  has been found; the stochastic problem is governed by a second-order Fokker-Planck equation, not a first-order continuity equation, and the Ott-Antonsen ansatz (\ref{eq:poisson}) no longer works in this case.  Perhaps there is some way to generalize the ansatz appropriately so as to reduce the stochastic model to a low-dimensional system, but for now this remains an open problem.  

\subsection{Directions for future research}
There are several other questions suggested by the work described here.  

\subsubsection{Validity of reduction method}\label{section6D1}
The most important open problem is to clarify the scope and limits of the Ott-Antonsen method used in Sec.~\ref{section2B}.  Under what conditions is it valid to assume that the infinite-dimensional Kuramoto model can be replaced by the low-dimensional dynamical system implied by the Ott-Antonsen ansatz?  Or to ask it another way, when do all the attractors of the infinite-dimensional system lie in the low-dimensional invariant manifold corresponding to this ansatz?  

This question has now become particularly pressing, because two counterexamples have recently come to light in which the Ott-Antonsen method~\cite{Ott-Ant} gives an incomplete account of the full system's dynamics.  When the method was applied to the problem of chimera states for two interacting populations of identical phase oscillators, it predicted only stationary and periodic chimeras \cite{Abrams}, whereas subsequent numerical experiments revealed that quasiperiodic chimeras can also exist and be stable \cite{Pikovsky-Rosenblum}.  Likewise, chaotic states are known to emerge from a wide class of initial conditions for series arrays of identical overdamped Josephson junctions coupled through a resistive load \cite{otherJJchaos, Watanabe}.   Yet the Ott-Antonsen ansatz cannot account for these chaotic states, because the reduced ODE system turns out to be only two dimensional \cite{Mirollo, Marvel}.  

What makes this all the more puzzling is that the method works so well in other cases.  It seems to give a full inventory of the attractors for the bimodal Kuramoto model studied here, as well as for the unimodal Kuramoto model in its original form~\cite{Kuramoto75, Kuramoto84, Ott-Ant} or with external periodic forcing~\cite{Ott-Ant, Antonsen}.    

So we are left in the unsatisfying position of not knowing when the method works, or why.  In some cases it (apparently) captures all the attractors, while in other cases it does not.  How does one make sense of all this? 

A possible clue is that in all the cases where the method has so far been successful, the individual oscillators were chosen to have randomly distributed frequencies; whereas in the cases where it failed, the oscillators were \emph{identical}.  Perhaps the mixing induced by frequency dispersion is somehow relevant here?  

A resolution of these issues may come from a new analytical approach.  Pikovsky and Rosenblum~\cite{Pikovsky-Rosenblum} and Mirollo, Marvel and Strogatz~\cite{Mirollo}  have independently shown how to place the Ott-Antonsen ansatz~\cite{Ott-Ant} in a more general mathematical framework by relating it to the group of Mobius transformations~\cite{Mirollo, goebel} or, equivalently, to a trigonometric transformation~\cite{Pikovsky-Rosenblum} originally introduced in the study of Josephson arrays~\cite{Watanabe}.   This approach includes the Ott-Antonsen ansatz as a special case, but is more powerful in the sense that it provably captures \emph{all} the dynamics of the full system, and it works for any $N$, not just in the infinite-$N$ limit.  The drawback is that the analysis becomes more complicated.  It remains to be seen what conclusions can be drawn---and, perhaps, what longstanding problems can be solved---when this new approach is unleashed on the Kuramoto model and its many relatives.   

Even in those instances where the Ott-Antonsen ansatz doesn't account for all the attractors of the full system, it can still provide useful information, for instance by giving at least some of the attractors and by easing the calculation of them.  Moreover, the transient evolution from initial conditions off the Ott-Antonsen invariant manifold can yield interesting phenomena not captured by the ansatz, as discussed in Appendix C of \cite{echo}.  

\subsubsection{Asymmetric bimodal distributions}
Now returning to the specific problem of the bimodal Kuramoto model: What happens if the humps in the bimodal distribution have unequal weights?  The analysis could proceed as in this paper, up to the point where we assumed symmetry between the two sub-populations. One would expect new phenomena such as traveling waves to arise because of the broken symmetry.  

\subsubsection{Finite-size effects}
We have focused here exclusively on the infinite-$N$ limit of the Kuramoto model.  What happens when the number of oscillators is reduced?  How do finite-size effects  influence the bifurcation diagram?  An analysis along the lines of~\cite{Hildebrand, Buice} could be fruitful for investigating these questions.

\subsubsection{Comparison with experiment}
Finally, it would be interesting to test some of these theoretical ideas in real systems.  One promising candidate is the electrochemical oscillator system studied by Hudson and colleagues \cite{Hudson}, in which the frequency distribution can be bimodal or even multimodal \cite{Mikhailov04}.

\section{Acknowledgments}
This research was supported in part by NSF grant DMS-0412757 and ONR award N00014-07-0734.

\appendix

\section{Alternative calculation of the boundary of stability for the incoherent state}
The system in Eq.~(\ref{eq:goveqns}), together with the bimodal natural frequency distribution given in Eq.~(\ref{bimodaldist}),
can be expressed using the formulation in \cite{Barreto} as two interacting populations of oscillators. In this case,
each population has a separate Lorenzian frequency distribution of width $\Delta$ and center
frequency at $\omega_0$ or $-\omega_0$, and the two-by-two matrix describing the relative coupling weights (i.e, Eq.~(1) in \cite{Barreto})
has $1/2$ in each entry. By postulating that a small perturbation to the incoherent state grows 
exponentially as $e^{st}$, and setting $s=i\nu$ for the marginally stable state,
Eq.~(9) of Ref.~\cite{Barreto} gives the following expression for the critical coupling value $K$:
\begin{equation}
K=\frac{2(\Delta^2-\nu^2+\omega^2_0)+i(4\Delta\nu)}{\Delta +i\nu}.
\label{a1}
\end{equation}
The boundary of stability of the incoherent state is obtained by requiring that this expression be strictly real.
One solution is obtained for $\nu=0$, resulting in $K=2(\Delta^2+\omega^2_0)/\Delta$, which is
equivalent to
\begin{equation}
\left(\frac{4\Delta}{K}-1\right)^2+\left(\frac{4\omega^2_0}{K}\right)^2=1.
\end{equation}
This is the equation for the semicircle in Figure \ref{fig:mainbifdiag1},
corresponding to a transcritical bifurcation of the incoherent state.
Another solution, obtained by assuming that $\nu \neq 0$ in Eq.~(\ref{a1}) and requiring $\omega_0 \geq \Delta$,
is $K=4\Delta$. This is the equation for the half-line in Figure \ref{fig:mainbifdiag1} corresponding to the
degenerate Hopf bifurcation of the incoherent state.

If the bimodal natural frequency distribution is given by a sum of Gaussians of
standard deviation $\sigma$ and centers at $\pm \omega_0$, then
the two-population approach outlined above leads to the following equation:
\begin{equation}
K=\sigma\sqrt{\frac{32}{\pi}}\left[F\left(\frac{\omega_0-\nu}{\sqrt{2}\sigma}\right)-F\left(\frac{-\omega_0-\nu}{\sqrt{2}\sigma}\right)\right]^{-1},
\end{equation}
where
\begin{equation}
F(z)=\frac{i}{\pi}\int^{\infty}_{-\infty}\frac{e^{-t^2}}{z-t}dt
\end{equation}
is known as the Faddeeva function and can be computed numerically \cite{Faddeeva}.
Once again requiring that $K$ be real, two branches corresponding to $\nu$ being
equal and not equal to zero can be obtained. These are the boundaries of stability
of the incoherent state shown in Fig.~\ref{fig:Gaussiandiag}.

\end{document}